\documentclass{PoS}

\usepackage{latexsym}
\usepackage{amsfonts}
\usepackage{amssymb}
\usepackage{amsmath}

\newcommand{\eq}[1]{\begin{align} #1 \end{align}}

\title{Multi moment cancellation of participant fluctuations}

\ShortTitle{MMCP method}

\author{\speaker{Viktor Begun}
 \\
        Warsaw University of Technology, Faculty of Physics, Koszykowa 75, 00-662 Warsaw Poland\\
        E-mail: \email{viktor.begun@gmail.com}}

\author{Maja Mackowiak-Pawlowska\\
        Warsaw University of Technology, Faculty of Physics, Koszykowa 75, 00-662 Warsaw Poland	\\
        E-mail: \email{majam@if.pw.edu.pl}}

\abstract{We summarize the new method for the correction of participant fluctuations in high energy nucleus-nucleus collisions. It allows to estimate a fluctuation baseline in comparison to a useful signal. In particular cases of a weak signal compared to baseline, it allows to cancel the baseline contribution from participants. }

\FullConference{Critical Point and Onset of Deconfinement - CPOD2017\\
		7-11 August, 2017\\
		The Wang Center, Stony Brook University, Stony Brook, NY}

\begin{document}

%
The way to probe the QCD phase diagram experimentally is to study interactions of various nuclei at different energies. One of the main background effects in such a study is fluctuations of participants - ${N_{\rm P}}$. This is the number of nucleons that interacted inelastically during a collision. Several popular ways of reducing participant fluctuations exist:
 \begin{enumerate}  
 \item the selection of narrow centrality bins, 
 \item the Centrality Bin Width Correction procedure~\cite{Luo:2011ts}, 
 \item the use of strongly intensive quantities~\cite{Gazdzicki:1992ri,Gorenstein:2011vq}.
\end{enumerate}
We checked the first two methods, and found that they still leave some participant fluctuations~\cite{Begun:2017sgs}. The third method requires an assumption that two measures e.g., pion and kaon multiplicities, are produced in the same volume and with the same volume fluctuations. It may be a too strong assumption.
We propose a different approach - to cancel participant fluctuations in a combination of several high fluctuation moments of a given quantity, e.g. particle type~\cite{Begun:2017sgs}. It may be possible, because fluctuations of participants give related contributions in low and high moments of measured distributions. It is similar to the third method, but we propose to use the moments from the same particle species. 
We call our new method - multi moment cancellation of participant fluctuations (MMCP).    

\paragraph{Fluctuation measures.}

\begin{figure}[h!]
\begin{center}
 \includegraphics[width=.7\textwidth]{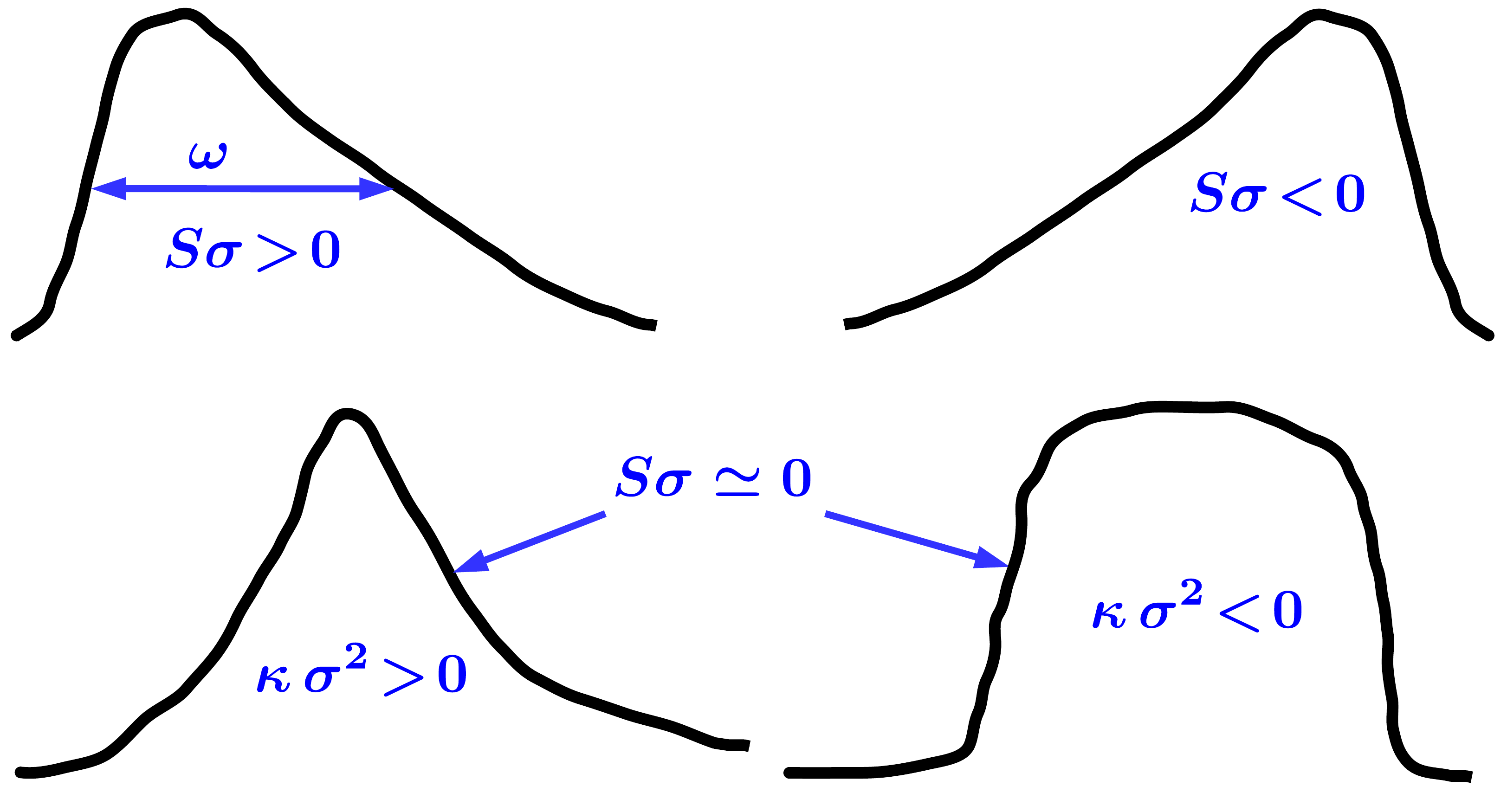}
 \caption{Relation of variance, skewness and kurtosis to the shape of a distribution. \label{forms}}
 \end{center}
\end{figure}

A multiplicity distribution, ${P(N)}$, can be characterized by central moments, ${m_n}$,
 \eq{{
 m_n ~=~ \sum_N\left(N-\langle N\rangle\right)^n P(N)~,}&&\text{where}\quad
 {
 \langle N^n\rangle ~=~ \sum_{N} N^n\, P(N)~.
 }}
which are related to cumulants, ${\kappa_n}$,
\eq{
 \kappa_2~=~m_2~,&& {\kappa_3~=~m_3~,}&& {\kappa_4~=~m_4~-~3m_2^2~,}&&{\ldots
 ~~.}}
%
%
Their special combinations  - {scaled variance}, normalized skewness, and normalized kurtosis,
 \eq{\label{w-Ss-Ks2}
   {{\omega}~=~\frac{\kappa_2}{\langle N\rangle} ~=~ \frac{\sigma^2}{\langle N\rangle}~,} &&
 {{S\,\sigma}~=~\frac{\kappa_3}{\kappa_2}~,} &&
 {{\kappa\,\sigma^2} ~=~ \frac{\kappa_4}{\kappa_2}~,}
 }
where $\sigma=\sqrt{\langle N^2\rangle-\langle N\rangle^2}=\sqrt{\kappa_2}$ is standard deviation, are used frequently, because they are related to the shape of a distribution, see Fig.~\ref{forms}
%
Note, that equivalence between Poisson and Gauss (Normal) distribution is broken already for $\omega$\footnote{For Poisson distribution $\omega=S\sigma=\kappa\sigma^2=1$, while for Gauss distribution $\omega$ is a free parameter, but $S\sigma=\kappa\sigma^2=0$. For another popular distribution - Log-Normal - one can select it's parameters so that $\omega\sim S\sigma\sim\langle N\rangle$ and $\kappa\sigma^2\sim\langle N\rangle^2$.  Therefore, the approach `just take Negative Binomial'  (Poisson, Gauss...) is not working for fluctuations.}. Therefore, selecting some baseline distribution imposes a certain relation between moments, which might not exist.
However, some assumptions must be done in order to proceed in analytical modeling. A possible minimal requirement is independent participant model. In this model the only assumptions are that participants are \textit{identical} and \textit{independent}. Then a mean multiplicity ${N}$ is the sum of contributions from ${N_{\rm P}}$ participants,
 \eq{
 {N~=~ n_1 ~+~ n_2 ~+~\ldots~+~ n_{_{N_{\rm P}}}~,}
 }
The {\textit{{identical}}} and {\textit{{independent}}} means that, ${\langle n_i \rangle = \langle n_j \rangle = \langle n_1 \rangle = \langle n_{\rm A}\rangle}$, and ${\langle n_i\,n_j\ldots n_k\rangle = \langle n_{\rm A}\rangle^k}$ . 
%
%
Then one can obtain~\cite{Begun:2016sop}:
 \eq{
 {\langle N\rangle} &~=~ {\langle N_{\rm P}\rangle}~ {\langle n_{\rm A}\rangle}~,
 \label{N1} \\
 \omega &~=~ \omega_{\rm A} ~+~ \langle n_{\rm A}\rangle ~ \omega_{\rm P}~,
 \label{w} \\
 {{S\,\sigma}} & {~=~ \frac{\omega_{\rm A}~{S_{\rm A}\,\sigma_{\rm A}}
 ~+~ \langle n_{\rm A}\rangle~{\omega_{\rm P}}\left[~3\,\omega_{\rm A} ~+~ \langle n_{\rm A}\rangle\,{S_{\rm P}\,\sigma_{\rm P}}~\right] }
        {\omega_{\rm A} ~+~ \langle n_{\rm A}\rangle ~ {\omega_{\rm P}}}~,}
  \label{Ss} \\
 {{\kappa\,\sigma^2}} & {~=~ \frac{\omega_{\rm A}~{\kappa_{\rm A}\,\sigma_{\rm A}^2}
        ~+~\langle n_{\rm A}\rangle~{\omega_{\rm P}}~
            \left[~\langle n_{\rm A}\rangle^2~{\kappa_{\rm P}\,\sigma_{\rm P}^2}
        ~+~\omega_{\rm A}
           \left(~
           3\,\omega_{\rm A}
        ~+~4\,S_{\rm A}\,\sigma_{\rm A}
        ~+~6\,\langle n_{\rm A}\rangle\,{S_{\rm P}\,\sigma_{\rm P}}
           ~\right)~\right]}
     {\omega_{\rm A} ~+~ \langle n_{\rm A}\rangle ~ {\omega_{\rm P}}}~,
   } \label{ks2}
 }
see also~\cite{Skokov:2012ds,Braun-Munzinger:2016yjz}. The values without index are those that can be measured. The A index labels the values that we would like to measure - fluctuations from one participant (a source). The P index labels participant fluctuations.
%
%
\paragraph{The MMCP method.}
Equations (\ref{N1}-\ref{ks2}) can be rewritten in a more compact form~\cite{Begun:2017sgs}: 
\eq{
 {\langle N\rangle} &~=~ {\langle N_{\rm P}\rangle}~ {\langle n_{\rm A}\rangle}~,
 \label{N11}\\
 {{\omega}} & {~=~ {\omega_{\rm A}}(1+{\alpha})~,} 
 \label{w1}\\
 {{S\,\sigma}} & {~=~ {\omega_{\rm A}}~\frac{3{\alpha}+{\delta}+{\alpha\,\beta}}{1+{\alpha}}~,}  
 \label{Ss1}\\
 {{\kappa\,\sigma^2}} & {~=~ {\omega_{\rm A}^2}~\frac{3{\alpha}+{\varepsilon}+{\alpha}\,\left[~6{\beta}+{\gamma}+4{\delta}~\right]}{1+{\alpha}}~.}
 \label{ks21}
}
The same as in Eqs.~(\ref{N1}-\ref{ks2}), on the left we have four values that can be measured - $\langle N\rangle$, $\omega$, $S\sigma$, $\kappa\sigma^2$, and eight unknowns on the right - $\alpha$, $\beta$, $\gamma$, $\delta$, $\varepsilon$, $\langle N_{\rm P}\rangle$, $\langle n_{\rm A}\rangle$, $\omega_{\rm A}$. However, in Eqs.~(\ref{N11}-\ref{ks21}) we introduced the values that characterize the strength of participant fluctuations compared to the fluctuations from a source,
 \eq{
 \alpha~=~\langle n_{\rm A}\rangle~\frac{\omega_{\rm P}}{\omega_{\rm A}}~, 
 &&
 \beta ~=~ {\langle n_{\rm A} \rangle}~\frac{{S_{\rm P}\,\sigma_{\rm P}}}{{\omega_{\rm A}}}~,
 &&
 \gamma~=~{\langle n_{\rm A}\rangle^2}~\frac{{\kappa_{\rm P}\,\sigma_{\rm P}^2}}{{\omega_{\rm A}^2}}~,
 }
and the values which characterize the relative strength of high and low order fluctuations from one source
 \eq{
 {{\delta}~=~\frac{{S_{\rm A}\,\sigma_{\rm A}}}{{\omega_{\rm A}}}} ~~~~\text{and}~~~~  {{\varepsilon}~=~\frac{{\kappa_{\rm A}\,\sigma_{\rm A}^2}}{{\omega_{\rm A}^2}}}
 }
It makes the analysis of Eqs.~(\ref{N1}-\ref{ks2}) easier. The relation,
 \eq{
 \alpha,~|\beta|,~|\gamma|~\ll~1~,
 }
is the mathematical meaning of the phrase `small participant fluctuations'.
For Gauss distribution $\varepsilon=\delta=0$, while for Poisson distribution  $\varepsilon=\delta=1$.
In a hadron gas, i.e. in a system away from critical point or phase transition, $\delta$ and $\varepsilon$ are in between Gauss and Poisson limit~\cite{Karsch:2010ck}
\eq{
 0~\leq~\delta = \varepsilon = \frac{1}{{\omega_{\rm A}^2}} = \tanh^2(\mu_B/T) ~\leq~1~, 
}
where $\mu_B$ is baryon chemical potential, and $T$ - temperature of the system.
Close to critical point $\kappa_n\sim\xi^{\frac{5(n-1)-1}{2}}$, where ${\xi\rightarrow\infty}$ is correlation length~\cite{Stephanov:2008qz,Mukherjeea:2017epo}, then
 \eq{
 {{\alpha} \sim {\langle n_{\rm A}\rangle^2}~\frac{{\omega_{\rm P}}}{\xi^2}\rightarrow 0~,} && 
 {{|\beta|} \sim {\langle n_{\rm A} \rangle^2}~\frac{{|S_{\rm P}\,\sigma_{\rm P}|}}{\xi^2}\rightarrow 0~, } 
 &&  
 {{|\gamma|} \sim {\langle n_{\rm A}\rangle^4}~\frac{{|\kappa_{\rm P}\,\sigma_{\rm P}^2|}}{\xi^4}\rightarrow 0~,} 
 \label{xi1}
 }
while
\eq{  
 {{\delta} \sim {\langle n_{\rm A}\rangle}~\xi^{0.5}\rightarrow \infty~,} 
 ~~~~~~\text{and}~~~~~~
 {{\varepsilon}\sim {\langle n_{\rm A}\rangle^2}~\xi\rightarrow \infty~.}
 \label{xi2}
 }
The non-observation of critical point suggests that $\xi$ is not large enough for the realization of approximations (\ref{xi1}) and (\ref{xi2}). Then, if they exist, the wanted `critical' fluctuations are comparable with the fluctuations of participants. 
For ${{\alpha},~{|\beta|},~{|\gamma|}\ll 1}$, and neglecting either ${{\delta}}$ or ${{\varepsilon}}$, one can solve the system (\ref{N11}-\ref{ks21}).
In case of  ${{\delta}\ll{\varepsilon}}$ there is a simple { analytic solution}:
\eq{
 {{\alpha}} &{~\simeq~ \frac{{S\,\sigma}}{3\,{\omega}-{S\,\sigma}}~,}
 \label{N111}\\
 {{\omega_{\rm A}}} &{~\simeq~ {\omega}~-~\frac{{S\,\sigma}}{3}~,}
 \\
 {{\kappa_{\rm A}\,\sigma_{\rm A}^2}} &{~\simeq~ {\kappa\,\sigma^2} ~-~{\omega_{\rm A}}\,{S\,\sigma}~,} \qquad {{\alpha},~{|\beta|},~{|\gamma|}~\ll~1,~~~{\delta}\ll {\varepsilon}~.}
 \label{ks211}
}
In this way one removes the fluctuations of participants and obtains the fluctuation from a source using {\textit{measured}} values.
%
\paragraph{Test of the MMCP.}

\begin{figure}[h!]
 \includegraphics[width=0.49\textwidth]{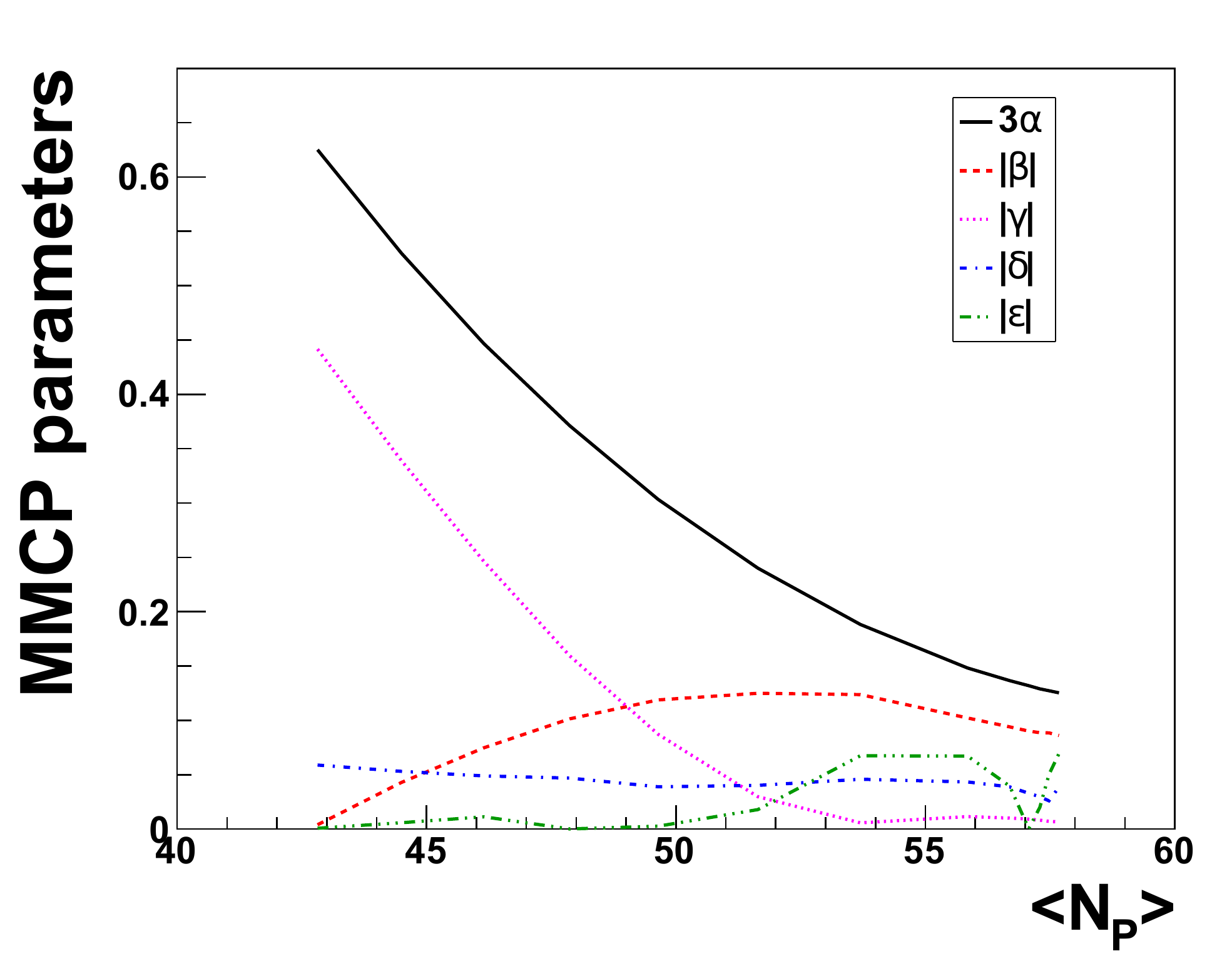}~~
 \includegraphics[width=0.49\textwidth]{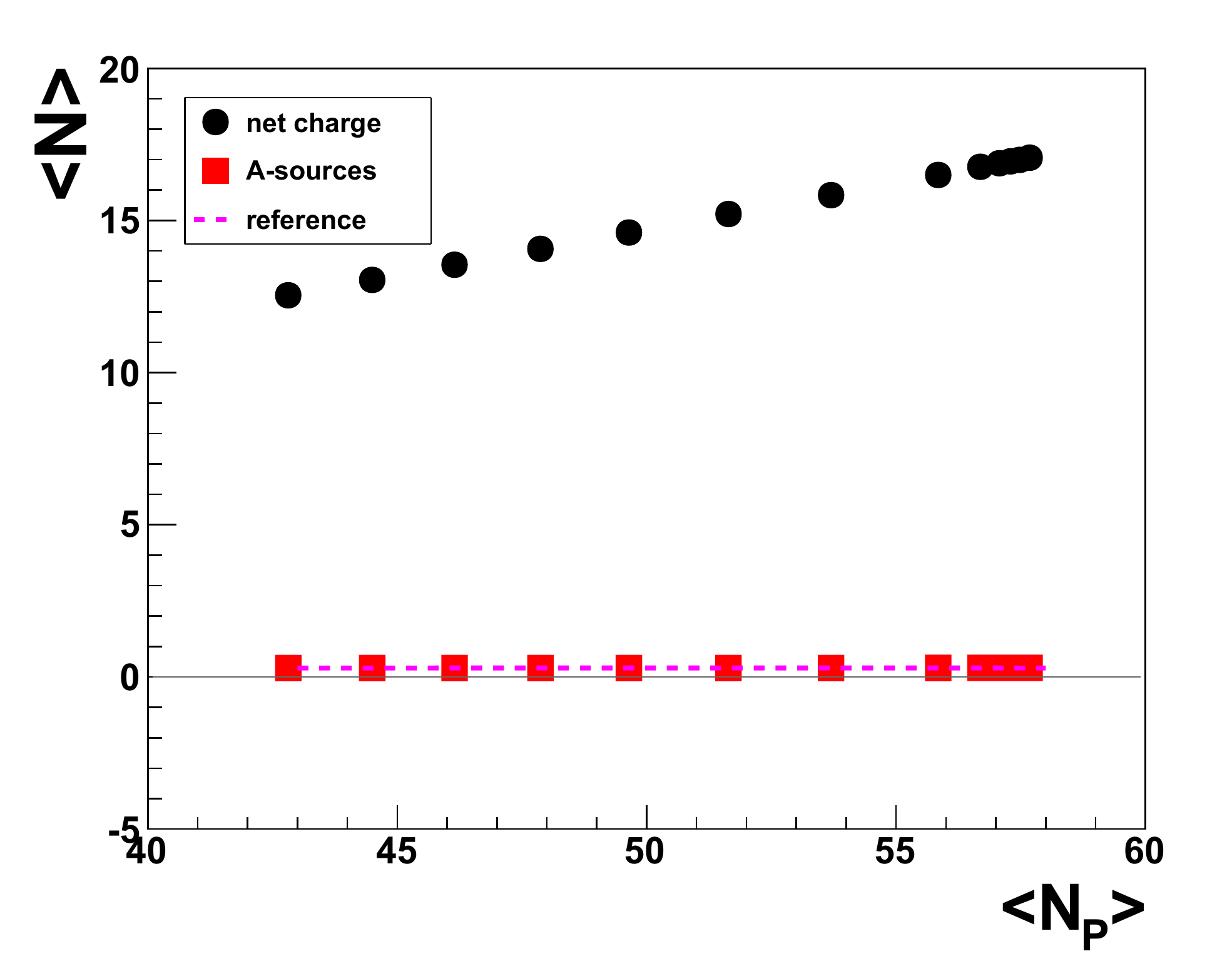}
 \caption{The dependence on centrality bin width for $\alpha$, $\beta$, $\gamma$, $\delta$, $\varepsilon$ (left), and $\langle N\rangle$, $\langle n_{\rm A}\rangle$ (right), see \cite{Begun:2017sgs}. 
 \label{fig-abg-N} }
\end{figure}

\begin{figure}[h!]
\begin{center}
 \includegraphics[width=0.5\textwidth]{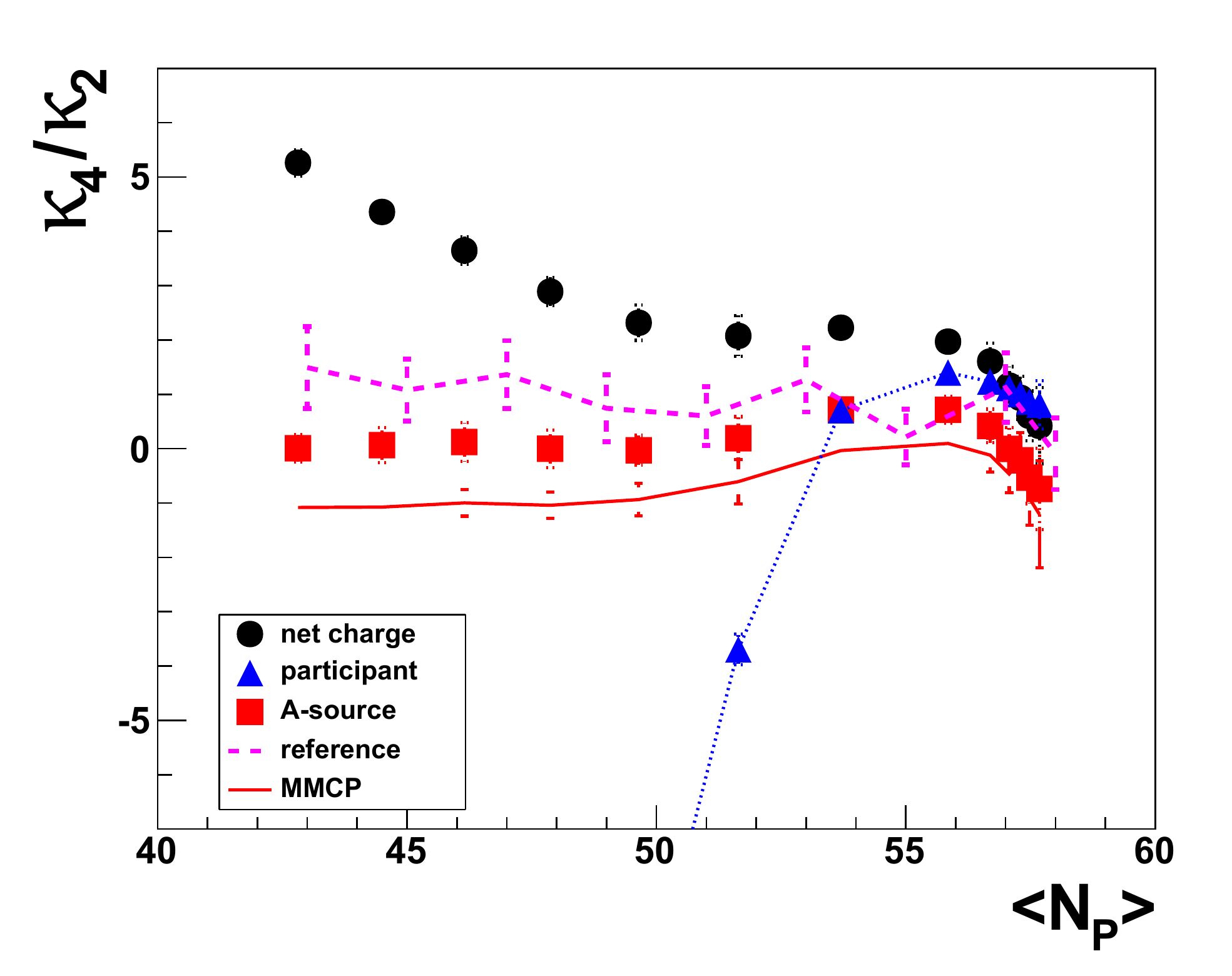}
 \caption{The same as Fig.~2 right for normalized kurtosis, see~\cite{Begun:2017sgs}. \label{fig-ks2}}
 \end{center}
\end{figure}

For the test of the above considerations we use the EPOS 1.99 model~\cite{Pierog:2009zt} applied to the net-electric charge in ${^{40}_{18}Ar+^{45}_{21}Sc}$ reactions at $ {p_{lab}=150}$~{ GeV/c} and with $ {y^{CMS}>0}$, see Figs.~\ref{fig-abg-N}-\ref{fig-ks2}.
The centrality selection is (left to right): $ {20\%}$, $17.5\%$, $15\%$, $12.5\%$, $10\%$, $7.5\%$, $5\%$, $2.5\%$, $1.5\%$, $1\%$, $0.75\%$, $0.5\%$ and ${0.2\%}$ with respect to zero centrality, i.e. from $0-20\%$ to $0-0.2\%$. The points correspond to the average number of participants in the corresponding centrality bins. The `reference' line is obtained for fixed $N_{\rm P}$, which is the closest integer number to a given $\langle N_{\rm P}\rangle$. 
One can see that for the system that we study the condition ${{\alpha},~{|\beta|},~{|\gamma|}~\ll~1}$ is valid.
The ${{|\delta|}~\ll~{|\varepsilon|}}$ is not always satisfied. However, a weaker condition is valid, ${{|\delta|},~{|\varepsilon|}<3{\alpha}}$, which is also enough for neglecting either $\delta$ or $\varepsilon$ in MMCP.
%
%
%
%
\begin{figure}[h!]
 \includegraphics[width=0.49\textwidth]{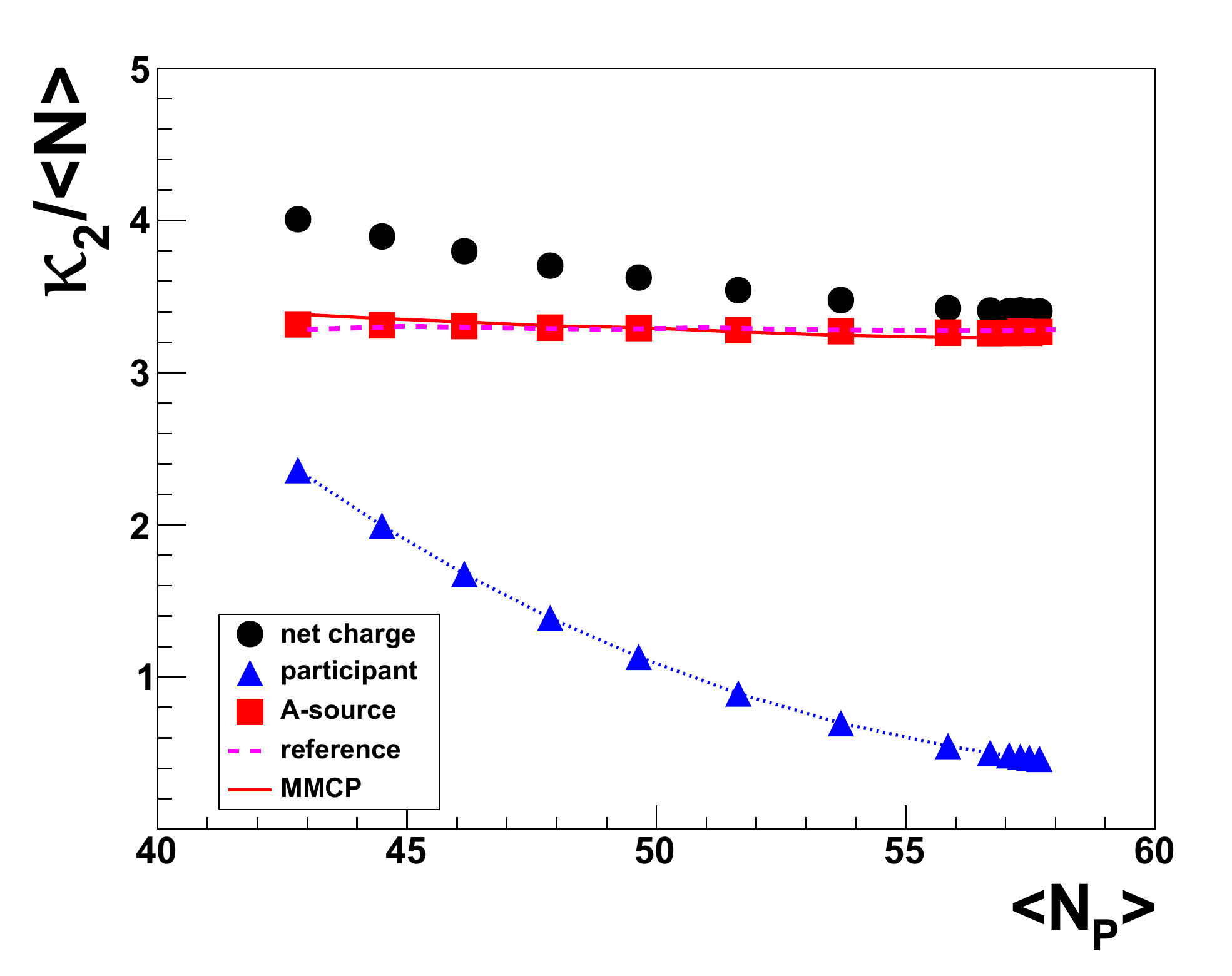}~~
 \includegraphics[width=0.49\textwidth]{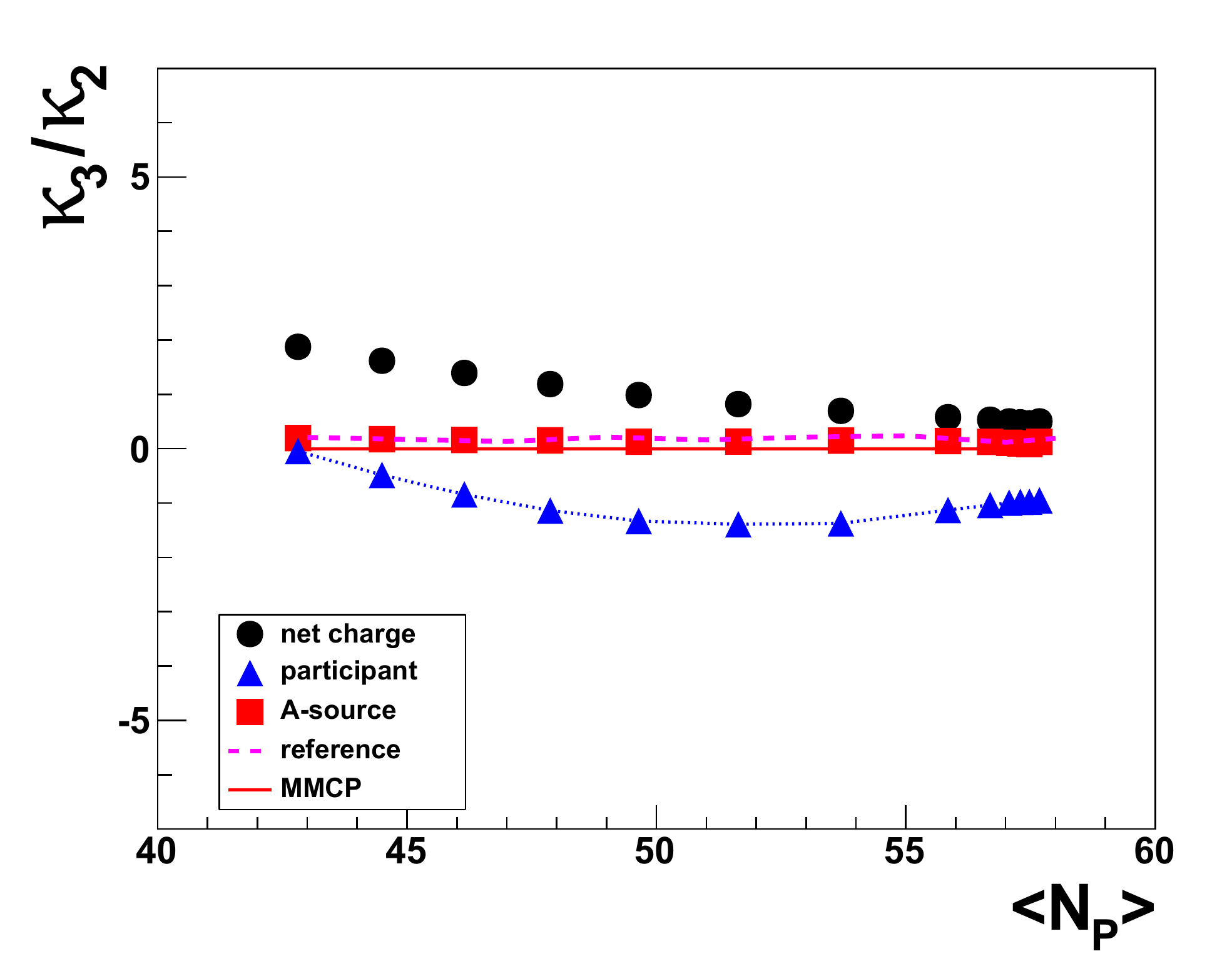}
 \caption{The same as Fig.~2 right for scaled variance and normalized skewness, see~\cite{Begun:2017sgs}.}
\end{figure}
%
%
%
%
%
%

%

%


\paragraph{Conclusions.} 

The multi moment cancellation of participant fluctuations procedure works very well for the scaled variance $\omega$. It reproduces a source fluctuations in all considered centrality bins. 
In case of normalized kurtosis, $\kappa\sigma^2$, the MMCP gives the values that are closer to a source, compared to the values which are obtained by narrowing centrality bins. 

The fluctuations from a source, ${S_{\rm A}\,\sigma_{\rm A}}$ and ${\kappa_{\rm A}\,\sigma_{\rm A}^2}$, { are almost zero} in the considered example. The largest contribution to the values that can be measured, ${S\,\sigma}$ and ${\kappa\,\sigma^2}$, gave lower order fluctuations of participants,
 \eq{{
 {S\,\sigma} ~\simeq~ 3\,{\langle n_{\rm A}\rangle}\, {\omega_{\rm P}}~>~0~, ~~~ {\kappa\,\sigma^2}~\simeq~3\,{\langle n_{\rm A}\rangle\,\omega_{\rm A}}\,{\omega_{\rm P}}~>~0~,}
 && \text{while}~~
  {{S_{\rm A}\,\sigma_{\rm A}} \simeq {0}~,~ {\kappa_{\rm A}\,\sigma_{\rm A}^2} \simeq {0}~.
 }}
Therefore, it is important to address participant fluctuations in details, at least for the third and the fourth fluctuation moment.
The average number of particles produced by a source, ${\langle n_{\rm A}\rangle}$, and it's fluctuations of the second, ${\omega_{\rm A}}$, and the third order, ${S_{\rm A}\,\sigma_{\rm A}}$, { do not depend on centrality}.
However, the fourth order fluctuations of a source, ${\kappa_{\rm A}\,\sigma_{\rm A}^2}$, change non-monotonously for the bin width smaller than 5\% in the range from $-1$ to $+1$. This  effect should be studied in more details.


%
%

\end{document}